\documentstyle[preprint,aps,eqsecnum,tighten]{revtex}
\begin{document}
\draft
\preprint{JIHIR 93--17}

\title   {
           Low--Energy $M1$ and $E3$ excitations \\
in the proton--rich Kr--Zr region
         }

\author  {       T. Nakatsukasa, K. Matsuyanagi
         }
\address {    Department of Physics, Kyoto University, Kyoto 606--01, Japan
         }

\author  {      I. Hamamoto\thanks{On
leave of absence from
Department of Mathematical Physics, Lund Institute of Technology,
Box 118, S-22100 Lund, Sweden},  W. Nazarewicz\thanks{On
leave of absence from
Institute of Theoretical Physics,
Warsaw University, Ho\.za 69, Warsaw, Poland;
Institute of Physics, Warsaw University of Technology,
Warsaw, Poland.}
         }
\address {           Joint Institute for Heavy-Ion Research,
                        Oak Ridge National Laboratory, \\
                   P.O. Box 2008, Oak Ridge, TN37831, U.S.A.\\
                                      and\\
                 Department of Physics, University of Tennessee,\\
                           Knoxville, TN 37996, U.S.A.
         }

\maketitle

\begin{abstract}
Low-energy intrinsic $K^\pi$=1$^+$, $0^-$, $1^-$, $2^-$, and $3^-$
states in the even-even
proton-rich Sr, Kr, and Zr nuclei are investigated using
the quasiparticle random phase approximation. In the Z$\simeq$N nuclei
the lowest-lying 1$^+$ states are found to carry unusually
large $B(M1)$ strength.
It is demonstrated that, unlike in the heavier nuclei,
the octupole collectivity
in the light zirconium region is
small and, thus, is not directly correlated with the
systematics of the lowest negative parity states.
\end{abstract}

\bigskip

\narrowtext
\section{Introduction}
It has been shown experimentally that shape coexistence, large
deformations, the presence of well-deformed intruder orbitals, quenching of
pairing correlations, low-lying octupole states, and dramatic shape changes
induced by rotation are quite common phenomena in the zirconium region
 (Z$\simeq$N$\simeq$40).
The microscopic reason for such a strong variation of collective properties is
the low single-particle level density in these medium-mass nuclei.
Because of spectacular shape effects, relatively small size,
and high collectivity,
the nuclei from the A$\sim$80 mass region
have become  favorite testing grounds for various theoretical
approaches.
Calculations based on the mean-field approach applied to nuclei in  the
light-Zr region
suggest an
interpretation of experimental data in terms of well-deformed prolate shapes,
weakly-deformed oblate shapes, and spherical (shell-model) configurations
\cite{[Naz85],[Ben88a]}.

There exist a number of mean-field calculations
for the light-Zr region [for references, see review \cite{[Woo92]}].
In most cases calculations give similar  equilibrium
deformations, but they differ in
their  predictions for excitation energies
of shape-coexisting states.
Best examples of the ground-state shape isomerism in nuclei in  the
light-Zr region are the Ge-Kr isotopes with A$\sim$70.
Calculations suggest the  interpretation in
terms of two competing configurations: one at an oblate shape, and
the other at a
prolate shape. Oblate ground states are predicted for
Ge- and Se-isotopes and for most Kr-isotopes.
For light Sr-isotopes the prolate configuration lies lower
in energy.
 Because of the mutual interaction
[of the order of a few hundred keV
\cite{[Ham82]}] the prolate and oblate bands are strongly
disturbed in the low-spin region.

The single particle diagram representative of the discussed nuclei
is shown in Fig.\ 1.
In the A$\sim$80 region both protons
and neutrons lie in the same ($p_{1/2}, p_{3/2}, f_{5/2}, g_{9/2}$) shell.
For $T_z$$\sim$0 systems, the proton and neutron shell
corrections  add coherently
and, consequently,  dramatic shape effects are expected.
A beautiful experimental signature of large prolate deformations
in the A$\sim$80 region,
attributed to the large single-particle gaps at Z,N=38 and 40,
 was observation of very collective rotational
bands in neutron-deficient Sr and Zr isotopes
\cite{[Lis82],[Pri83]}.

The investigation of the medium-mass N=Z nuclei has been the proprietary
niche of groups who made investigations using  the Daresbury Recoil Separator.
Pioneering works
from Daresbury include the spectroscopy of
$^{64}$Ge,
$^{68}$Se, $^{72}$Kr,
$^{76}$Sr,
$^{80}$Zr,
and
$^{84}$Mo [see ref. \cite{[Lis90a]}].
These studies confirmed earlier theoretical predictions of shape
transition from strongly oblate shapes in
$^{68}$Se and  $^{72}$Kr
to strongly  prolate shapes in
$^{76}$Sr,
and $^{80}$Zr (actually,
$^{76}$Sr and $^{80}$Zr are,
according to calculations,
very deformed, with the ground state deformation
around $\beta_2$=0.4).
The nucleus $^{84}$Mo is the heaviest Z=N system known so far.

Spectroscopy in the light-Zr region
will certainly become one of the main arenas
of investigations  around the proton drip line.
The physics of exotic nuclei
with $T_z$${\; \raisebox{-0.4ex}{\tiny$\stackrel
{{\textstyle<}}{\sim}$}\;}$0
is one of the fastest developing subjects
in nuclear physics,
thanks to exotic (radioactive) ion beam ({\sc rib}) facilities
currently under construction in Europe, U.S.A., and Japan.
In particular, the combination of {\sc rib} and
the new-generation multidetector arrays should open up many new
avenues of exploration \cite{Dourdan}.

The main motivation of this paper is to make predictions
for low-energy collective $M1$ and $E3$ excitations around $^{76}$Sr.
Since the $M1$ collectivity of low-lying $1^+$ states increases
with deformation (though the energies of those states may increase),
it is anticipated that in some well deformed nuclei in the
A$\sim$80 mass region the strong
magnetic dipole strength should lie
low in energy.
The existence of collective octupole states
in this region is a long-standing question. The low-lying negative-parity
states, often interpreted as octupole vibrations, can be of
a single-particle character \cite{[Enn92]}. To shed some light on
both issues we performed calculations
based on the quasiparticle Random Phase Approximation
({\sc rpa}). We hope, that those
predictions will stimulate experimental investigations of medium-mass
nuclei around
the N=Z line.

\section{Deformations and
Pairing Correlations in the A$\sim$80 Mass Region}\label{pairing}

Calculations of equilibrium deformations of A$\sim$80
isotopes were previously performed \cite{[Naz85]}
within the Woods-Saxon-Strutinsky model
\cite{[Cwi87]}. In this work,    new calculations
have been carried out using  the
same single-particle model but the
  Yukawa-plus-exponential  mass  formula  of
ref.\ \cite{[Mol88]}. The
particle-particle interaction
 was approximated by  the state-independent
monopole-pairing Hamiltonian.
The pairing energy was computed using the
approximate particle number projection
in the   Lipkin-Nogami version.
The pairing strengths and the
 average pairing energy were taken according to ref.\ \cite{[Mol92b]}.
The calculated equilibrium deformations
for selected Kr, Sr, and Zr isotopes
are shown in Table~\ref{defs}.
It is seen that the deformed$\rightarrow$spherical shape transition
is expected to occur around N$\sim$44. Worth noting are very
large equilibrium $\beta_2$ deformations ($\sim$0.4)
of the lightest Kr, Sr, and Zr isotopes.

In several nuclei around $^{82}$Sr highly-deformed and superdeformed
bands ($\beta_2$$>$0.4) have been predicted
to become yrast at high spin \cite{[Naz85],[Naz88],[Rag88],[Dud87a]}.
For example, in  $^{82}$Sr  well deformed nearly-prolate
bands involving $h_{11/2}$ neutrons
are expected to become yrast
at I$>$32$\hbar$.  Experimentally, a weak ridge-valley structure with a width
of $\Delta E_\gamma$$\approx$150 keV has been seen in the
$E_\gamma$--$E_\gamma$
correlation map \cite{[Bak91]}. This ridge corresponds to
$\beta_2$$\sim$0.5 for a deformed
rigid rotor. However, no discrete band that could be associated with this
ridge-valley was identified so far.
Theoretically, the superdeformed band in $^{82}$Sr is expected
\cite{[Naz85]} to have deformation $\beta_2$$\sim$0.45, see Table~\ref{defs}.

The most important interaction,
beyond the single-particle deformed mean
field, is the short-ranged pairing interaction. This force is often
approximated by means of a state-independent monopole pairing interaction.
The general feature of the pairing interaction is that
the pair correlation energy
is anticorrelated with the shell correction.
A smaller pairing gap results from a smaller density of single-particle
levels around the Fermi level, which are available for pair correlation.
For deformed A$\sim$80 nuclei the weakest pairing is expected around
the deformed gaps at N (or Z)=38--42 \cite{[Naz88]}.
A further reduction of pairing can occur in excited
configurations, due to blocking.

In the A$\sim$80 mass region are several good examples of
very regular, rigid rotational bands.
Among them there are negative parity
bands in $^{76}$Kr and $^{78}$Kr built upon the first $I^\pi$=3$^-$ state
at 2258 keV and 2399 keV, respectively. These bands
are among the best  {\em normally-deformed} rotors,  with remarkably large
and nearly constant moments of inertia,
${\cal J}^{(1)}$$\approx$${\cal J}^{(2)}$ \cite{[Kap88],[Gro89]}.
Theoretically, those bands are associated with two-quasiparticle
excitations built upon the proton [431 3/2]$\otimes$[312 3/2] Nilsson
orbitals which happen to occur just below the strongly deformed
subshell closure at Z=38.
[The proton character of those bands was recently confirmed
by the $g$-factor measurement \cite{[Bil93]}.]
 Another good example is the [312 3/2] band
in $^{77}$Rb  \cite{[Luh86]}
or the [422 5/2] band in $^{81}$Y \cite{[Lis85]}
having unusually large moments of inertia. In all those cases the
{\sc bcs} calculations \cite{[Naz88]}
 suggest the dramatic reduction (or collapse)
of the static pairing.

Weak pairing has important consequences for the low-energy
electromagnetic transitions.
Since the $B(M1)$
values involving the ground state of even-even nuclei
are proportional to the {\sc bcs}
factor $(u_\mu v_\nu-v_\mu u_\nu)^2$,
 weaker pair correlations
enhance the low-lying $M1$ strength.
For electric  transitions, the related
{\sc bcs} factor is  $(u_\mu v_\nu+v_\mu u_\nu)^2$.
On the average, pairing correlations enhance the collectivity
of the low-lying $E3$ transitions
from/to the ground state in
 the Sr-Zr region (see Sec. \ref{EE3}).

\section{Magnetic Dipole States}\label{M1}

The deformation dependence of $1^+$ states is a current subject of both
experimental \cite{[Zie90],[Mar93]}  and theoretical
\cite{[Bes88],[Ham92]} studies.
The low-energy $B(M1)$ strength (defined as
the summed strength over a given energy interval,
e.g., 2--4 MeV in the rare-earth nuclei)
increases with quadrupole deformation as, roughly, $\beta_2^2$.
Recently, it was demonstrated  in ref. \cite{[Ham92]} that
the sum of $B(M1)$ values in the region
of $E_x$$<$10 MeV at heavy
superdeformed nuclei around $^{152}$Dy and $^{192}$Hg
was several times larger than
that at normal deformations.
The reason for this enhancement is twofold. Firstly, the
proton convection current contribution
to $B(M1)$ increases with deformation
and at strongly deformed shapes becomes comparable to the spin-flip
contribution
in the low-energy region. Secondly, as discussed in Sec. \ref{pairing},
the $B(M1)$
strength increases if
the pair correlations are weak, i.e., exactly what is expected at
 {\sc sd} shapes \cite{[Naz91a]}.

Since some of the A$\sim$80 nuclei are very well deformed in their
ground states,
their equilibrium deformations
exhibit rapid isotopic and isotonic variations,
and
their pairing correlations are
predicted to be weak due to deformed subshell closures
(Table~\ref{defs}). Because
the Kr, Sr, and Zr isotopes
have these characteristics, they
are ideally suited for investigations of the
low-energy $M1$ strength and its
deformation dependence. (The lighter and heavier systems,
such as Ge, Se,  and Mo, are less deformed and $\gamma$-soft.)

The properties of the $K^\pi$=1$^+$ states
have been
investigated using the
{\sc rpa} Hamiltonian
\begin{equation}\label{HRPA}
H_{QRPA}=h_{s.p.}+V_{pair}+V_{FF}+V_{\sigma\sigma},
\end{equation}
where the single-particle Hamiltonian,
\begin{equation}
h_{s.p.}=\sum_{i}(\epsilon_i-\lambda)c_i^\dagger c_{i}
\end{equation}
 is an axially deformed Woods-Saxon Hamiltonian of
ref.\ \cite{[Ehr72]} [see ref.\ \cite{[Dud81]} for parameters],
\begin{equation}\label{Vpair}
V_{pair}=-\Delta \sum_i (c_i^\dagger c_{\bar i}^\dagger
+c_{\bar i}c_i)
\end{equation}
is the monopole-pairing field,
 $V_{FF}$ is a long-ranged
residual interaction (mainly of qua\-dru\-po\-le-quadrupole type),
and $V_{\sigma\sigma}$ is the spin-spin residual interaction.
In eq.~(\ref{HRPA})
\begin{equation}\label{VF}
V_{FF}=-{1\over 2}\sum_{T=0,1}\kappa_T F_T^+F_T,
\end{equation}
where the isoscalar and isovector fields $F$ are given by
\begin{equation}\label{VF1}
F_{T=0}=F_n+F_p\,\ ;~~~F_{T=1}=F_n-\xi F_p
\end{equation}
and
\begin{equation}
F_\tau ={1\over{i\hbar}}\left[h_{s.p.}^{(\tau)},
j_+^{(\tau)}\right],~~~\tau=n,p
   ,
\end{equation}
while the residual spin-spin interaction
is written as
\begin{equation}\label{VS}
V_{\sigma\sigma}={1\over 2}\sum_{T=0,1}\chi_T S_T^+S_T,
\end{equation}
where
\begin{equation}\label{VS1}
S_{T=0}=S_n+S_p\ ;~~~S_{T=1}=S_n-S_p.
\end{equation}
The strength of $V_{\sigma\sigma}$
is taken
\cite{[Ham84]} as $\chi_0$=$\chi_1$=100/A\,MeV.

     The residual interaction $V_{FF}$
 gives rise to
isoscalar and isovector shape oscillations. The isoscalar-coupling constant,
$\kappa_0$, is determined by the condition \cite{[Ham71]}
that the lowest {\sc rpa} frequency for
the isoscalar mode vanishes, since the lowest-lying mode with
$K^\pi$=1$^+$ is
spurious
and corresponds to a uniform rotation of the system.
The value of  $\xi$  in (\ref{VF1}) is determined by the
requirement \cite{[Pya72]} that the spurious component should be
 absent in the {\sc rpa}
solutions with non-zero frequencies.
We have numerically checked that
the summed probability of the spurious component,
$|S\rangle$ $\propto$ $j_+|g.s.\rangle$,
remaining in
the {\sc rpa} solutions with non-zero frequency is less than
10$^{-6}$.

The isovector coupling constant, $\kappa_1$,  is taken from the
self-consistency condition for the harmonic oscillator model \cite{[Boh75]},
$\kappa_1$=--3.5$\kappa_0$.
In {\sc rpa} calculations
we take
into account all two-quasiparticle configurations with excitation energies less
than 26  MeV, and
have checked that the configuration space is sufficiently large so
as to include all $M1$ strengths.

As a representative example, results of calculations for Sr isotopes
are shown in Fig.\ 2,
which shows the excitation energies of the low-lying $K^\pi$=1$^+$ states.
The values $B(M1;g.s\rightarrow 1^+)$
(in $\mu_N^2$) are indicated.
 The upper diagram was obtained by using
the standard  pairing gaps of Table~\ref{defs}.
According to Sec.~\ref{pairing}, pairing correlations in the
excited states of Sr-Zr are expected to be seriously quenched. Therefore,
we performed a second set of calculations with $\Delta_p$
and $\Delta_n$ reduced by  50\% with respect to the standard values.
As discussed in refs.\ \cite{[Ham91],[Ham92]}, reduced pairing leads to
increased collectivity of the low-lying 1$^+$ states; as
seen in Fig.\ 2
the $B(M1)$ values calculated in the ``weak pairing" variant are
approximately twice as
large as the $M1$ rates obtained in the ``standard pairing" variant.

The best candidate for low-lying enhanced 1$^+$ states in the A$\sim$80
mass region is the N=Z nucleus $^{76}$Sr. Its ground state is very well
deformed due to the coherent superposition of proton and neutron shell effects
associated with the deformed gap at the particle number 38.
     In Fig.\ 3
we show the  $B(M1;g.s.\rightarrow 1^+)$ strengths
of the calculated $K^\pi$=1$^+$ {\sc rpa}
excitation modes in $^{76}$Sr
(at  the ground-state
 deformation), as a function
of  excitation energy.
The upper (lower)  diagram corresponds to the standard (weak) pairing variant.
The $M1$ strength arising from only the proton convection current
(i.e., $g_s$=0)
and the $M1$ strength  from only the spin part
(i.e., $g_l$=0)
are also plotted
in Fig.\ 3.
In both
pairing variants of calculations, there appears only one
low-lying 1$^+$ state which has unusually strong
 $M1$ collectivity. In the
``weak pairing" variant this state is predicted at 2.2 MeV and the
corresponding $B(M1;g.s.\rightarrow 1^-)$ transition is  2.16 $\mu_N^2$.
The main components of the
wave function of the 1$^+$ state in $^{76}$Sr
are the $\pi(g_{9/2})^2$ and  $\nu(g_{9/2})^2$ excitations
involving the two Nilsson orbitals [431 3/2] and
[422 5/2].
The largest components of
the low-lying $1^+$ states in $^{76}$Sr
in the energy range of 4--5 MeV  are
the [431 3/2]$\otimes$[431 1/2] (spin-flip) and
[301 3/2]$\otimes$[310 1/2] two-quasiparticle excitations.
The main contribution to the peak in the $M1$ distribution
seen in the energy range of 7--9 MeV in Fig.\ 3 comes
almost exclusively from  the spin-flip $f_{7/2}\rightarrow f_{5/2}$
and $g_{9/2}\rightarrow g_{7/2}$ transitions.

The contribution to the $B(M1)$ strength coming from the
unique-parity
high-$j$ excitations, such as $(h_{11/2})^2$ or $(g_{9/2})^2$,
has a simple shell model interpretation (in terms of a single-$j$
shell)
and cannot be viewed as
coming from  a collective ``scissors" mode [see discussion in
ref. \cite{[Ham84]}].
The synthetic orbital scissors state is defined as
\begin{equation}\label{scissor}
|R\rangle = {\cal N}^{-1}\left(l_+^{(n)}-\alpha l_+^{(p)}\right)|g.s.\rangle,
\end{equation}
where ${\cal N}$ is a normalization factor and the parameter $\alpha$ is
determined by the requirement that the mode (\ref{scissor})
is orthogonal to the spurious reorientation mode
\cite{[Ham92],[Fae90],[Zaw91]}, i.e.,
\begin{equation}\label{alpha}
\alpha=\langle g.s.|j_-^{(n)} l_+^{(n)}|g.s.\rangle /
\langle g.s.|j_-^{(p)} l_+^{(p)}|g.s.\rangle.
\end{equation}
The calculations show that for the
lowest 1$^+$ state in $^{76}$Sr the
overlap between its {\sc rpa} wave function and
the state (\ref{scissor}) is only about 12\%.
Consequently, although this state is predicted to carry
an unprecedented $M1$ strength,
it cannot
be given a geometric interpretation of the ``scissors" mode.
The $K^\pi$=1$^+$ isovector
giant quadrupole resonance
in $^{76}$Sr lying
at $E_{ex}$$\sim$32 MeV
carries a significant  $M1$
 strength ($\sim$4 $\mu_N^2$) and contains a major component of
the ``scissors mode" (around 50\%).

Figures\ 4 and 5 show the calculated 1$^+$ states
in Kr and Zr isotopes, respectively.
As seen in Figs.\ 2, 4, and 5
when moving away from $^{76}$Sr, the low-energy $M1$ strength becomes
more fragmented.
Good prospects where to find
large $M1$ strength at low energies are the well-deformed prolate nuclei
$^{78}$Sr (where the 1$^+$ state is built mainly from the
$\pi$([431 3/2]$\otimes$[422 5/2]) and
$\nu$([422 5/2]$\otimes$[413 7/2])
 two-quasiparticle excitations),
$^{80}$Sr,
$^{80}$Zr
($\pi$([422 5/2]$\otimes$[413 7/2]) and
$\nu$([422 5/2]$\otimes$[413 7/2])),
 $^{82}$Zr, and $^{74}$Kr.
The most promising oblate-shape candidate
is the N=Z nucleus
$^{72}$Kr. Similar to $^{76}$Sr,  the  $1^+$ state
in $^{72}$Kr has a $(g_{9/2})^2$ character. However, in this case
the main contribution comes from the high-$\Omega$ substates, i.e.,
$\pi$([413 7/2]$\otimes$[404 9/2]) and
$\nu$([413 7/2]$\otimes$[404 9/2]).

As discussed in Sec.~\ref{pairing}, the best prospects for superdeformation in
 the A$\sim$80
region are in the nuclei around $^{82}$Sr. The calculations
performed for superdeformed configuration of $^{82}$Sr
 predict two states (around 3 MeV and 4 MeV) that carry a large $M1$ strength
(see Fig.\ 2).
They can be associated with the
$\pi$([431 3/2]$\otimes$[422 5/2]),
$\nu$([422 5/2]$\otimes$[413 7/2]) and
$\nu$([541 3/2]$\otimes$[550 1/2]) two-quasiparticle excitations.

\section{Octupole Correlations}\label{EE3}

In the light zirconium
region octupole correlations can be associated with the $g_{9/2}$ and
$p_{3/2}$ subshells. Because of their rather large energy separation
and
a small
number of coupling matrix elements, no pronounced octupole
instability is
expected. In addition, the small number  of  active
subshells makes the octupole effect more sensitive to quadrupole
distortion than
in heavier nuclei around $^{146}$Ba or $^{222}$Th \cite{[Naz90b]}.

The systematics of the lowest $3^-$
excitations in the Zr-region is shown  in
Fig.\ 6.
It is seen that
$E_{3^-}$ tends to decrease
when approaching the nucleus $^{76}$Sr.
On the other hand,
the shell correction calculations \cite{[Enn92],[Naz84],[Ska91]}
 predict octupole
softness only in the transitional isotopes of Zn-Se with N$\le$36.
Is the presence of
low-lying negative-parity state always a good fingerprint of
octupole collectivity? The answer to this question is negative.
There are many nuclei that possess relatively high-lying negative
parity excitations but still are considered as good examples of
systems with strong octupole correlations.
In fact,
the systematics of experimental $B(E3)$
values in the light-Zr region \cite{[Enn92],[Spe89]}
indicates that no correlation can be found between
the behavior of the
lowest negative-parity states shown in Fig.\ 6 and
the $B(E3;g.s. \rightarrow 3^-)$ strength.

According to the energy systematics presented in Fig. 6,
the lowest negative-parity  states are observed in
strongly deformed nuclei with
particle number (N or Z) close to 38. For example, in the nucleus
$^{76}$Kr two negative-parity rotational bands built upon the
(3$^-$) (2258 keV)
and (2$^-$) (2227 keV) bandheads are known.
However,
the coexisting prolate and oblate  minima
in this nucleus are predicted \cite{[Enn92]}
to be  fairly rigid with respect
to the reflection-asymmetric distortion.
In ref.\ \cite{[Cot89]}, based on
energy systematics,  it has been argued that some negative parity bands
in well-deformed nuclei from
the A$\sim$80 mass region can  be interpreted as
collective (aligned) octupole
bands. However, it is not the excitation energy of the negative parity
band itself that determines the collective character of the underlying
intrinsic configuration.
In $\pi$=-- bands pairing correlations are usually reduced due to blocking
and there is also significant Coriolis mixing. Consequently, these
bands have usually larger moments of inertia
than ground bands and, in some cases, can become yrast at high spins.
In our opinion, the observed lowering of negative-parity states around
the particle number 38 does not
necessarily indicate strong octupole
correlations as suggested in ref.\ \cite{[Cot89]}  but rather has
a non-collective origin, see below.

In order to clarify the issue of octupole collectivity around
Z=38, N=38 we performed the {\sc rpa} calculations
with the Hamiltonian
\begin{eqnarray}
H_{QRPA}
 &=& h_{s.p.} + V_{pair}
-{1\over 2}\sum_K \chi_{3K}^{T=0} Q^{''\dagger}_{3K}Q^{''}_{3K}
-{1\over 2}\sum_K \chi_{3K}^{T=1}
(\tau_3Q_{3K})^{''\dagger}(\tau_3Q_{3K})^{''}\nonumber \\
&+& {1\over 2}\sum_K \chi_{1K}^{T=1}
(\tau_3D_{1K})^{''\dagger}(\tau_3D_{1K})^{''} \ .
\label{hamiltonian}
\end{eqnarray}
where $h_{s.p.}$ is a single-particle Nilsson Hamiltonian,
$V_{pair}$ is given by (\ref{Vpair}),
and $Q_{3K}^{''}=(r^3Y_{3K})^{''}$ [$D_{1K}^{''}=(rY_{1K})^{''}$] are the
doubly-stretched octupole (dipole) operators \cite{[Sak89]}.
A large configuration  space
composed  of  7 major shells (for both protons  and  neutrons)
was used  when
solving the coupled {\sc rpa} equations.
The octupole isoscalar coupling
 strengths, $\chi_{3K}^{T=0}$, were determined by the
self-consistency  condition
for the harmonic oscillator model
\cite{[Boh75],[Sak89]},
\begin{eqnarray}
\chi _{3K}^{T=0} &=& {4\pi  \over 7} M\omega ^2_0 \left\{
\langle (r^4)^{''} \rangle _0 + {2 \over 7} (4 - K^2)
\langle  (r^4P_2)^{''} \rangle  _0 \right.\nonumber \\
 &+& {1 \over 84} \left.\left[K^2 (7K^2 -67) + 72\right]
\langle  (r^4P_4)^{''} \rangle  _0 \right\} ^{-1} \ .
\label{forcestrength}
\end{eqnarray}
The strength of the isovector octupole mode was taken from ref.
\cite{[Vej66]}
\begin{equation}
\chi_{3K}^{T=1} = -0.5\chi_{3K}^{T=0},
\end{equation}
while for the isovector dipole mode we used the value
\cite{[Boh75],[Sak89]},
\begin{equation}
\chi_{1K}^{T=1} = \frac{\pi V_1}{\langle (r^2)^{''}\rangle}M\omega_0^2
\end{equation}
with $V_1$=140MeV.
A similar model has been used recently \cite{[Miz90],[Miz91]} to discuss
octupole excitations built upon superdeformed shapes.

It is worth noting that,
because we use the doubly-stretched
$Q^{''\dagger}_{3K} Q^{''}_{3K}$ interactions,
there is no simple correlation between the number of
two-quasiparticle configurations contributing to an excited state
and the corresponding
 $B(E3)$ value. That is, an excitation which looks fairly collective
in terms of the {\sc rpa} amplitudes
(i.e.,
appreciable size of backward-going amplitudes),
it still can have a very small
$B(E3)$ value.
Indeed,
the ordinary octupole strengths $|\langle n |Q_{3K}|0\rangle |^2$ are quite
different from the doubly-stretched octupole strengths
$|\langle n |Q_{3K}^{''}|0\rangle |^2$ in well-deformed nuclei.
For example, in case of the prolate superdeformed harmonic oscillator potential
($\omega_\bot = 2\omega_3$),
ratios of the energy-weighted sum rule
 values $S_{3K}$ (for $Q_{3K}$ operators) and $S_{3K}^{''}$
(for $Q_{3K}^{''}$ operators) are given by \cite{[Nak92]}
\begin{equation}\label{SR1}
S_{3K} : S_{3K}^{''}=\left\{
\begin{array}{rl}
50 : 11, &\quad\mbox{for $K=0$,}\\
13 : 4, &\quad\mbox{for $K=1$,}\\
1 : 1, &\quad\mbox{for $K=2$,}\\
1 : 4, &\quad\mbox{for $K=3$,}\\
\end{array} \right.
\end{equation}
while in the oblate superdeformed case ($\omega_3=2\omega_\bot$),
\begin{equation}\label{SR2}
S_{3K} : S_{3K}^{''}=\left\{
\begin{array}{rl}
5 : 8, &\quad\mbox{for $K=0$,}\\
17 : 26, &\quad\mbox{for $K=1$,}\\
1 : 1, &\quad\mbox{for $K=2$,}\\
4 : 1, &\quad\mbox{for $K=3$.}\\
\end{array} \right.
\end{equation}
Therefore, in the well-deformed
prolate (oblate) configurations, $B(E3)$ values
overestimate (underestimate) the
collectivity  (in the sense of the {\sc rpa} with doubly-stretched
interaction) for the $K^\pi$=0$^-$ and
1$^-$ states, while they underestimate (overestimate)
the ``doubly-stretched" octupole
collectivity of the $K^\pi$=3$^-$ states.

The results of calculations for the Sr isotopes are shown in Fig.\ 7,
which displays
the predicted
excitation energies of intrinsic $K^\pi$=0$^-$, 1$^-$, 2$^-$,
and 3$^-$ states and the corresponding $B(E3)$ values
(in s.p.u.). The forward {\sc rpa} amplitudes for the
$0^-$, $1^-$, $2^-$,
  and 3$^-$ states built upon  prolate configurations in $^{76,78,80,82}$Sr
are plotted in Figs.\ 8-11, respectively.
In none of the nuclei considered, the
low-lying negative-parity excitations
can be considered as highly-collective states.

In the N=Z nucleus $^{76}$Sr the lowest negative-parity
excitations with $K$=1 and 2 can be considered as weakly collective.
The $K$=1 octupole phonon has a large component
of the two-quasiparticle [312 3/2]$\otimes$[422 5/2]
neutron configuration,
see Fig. 9. The $K^\pi$=2$^-$  mode  is less collective
but it lies lower in energy.
As seen in Fig. 10, the main contribution to its wave
function comes from the [310 1/2]$\otimes$[422 5/2] proton and neutron
excitations. The lowest $K^\pi$=0$^-$ excitation is mainly built
upon the [312 3/2]$\otimes$[431 3/2] excitations.
The $K^\pi$=3$^-$ state is  predicted to be a non-collective
[310 1/2]$\otimes$[422 5/2] state,
see Fig.\ 11.
Of course, all
those intrinsic states are expected to be mixed through the Coriolis
interaction \cite{[Ner70]}.
In the ``weaker pairing" variant of the calculations, the
$B(E3;g.s.\rightarrow 1^-)$ rate is reduced by a factor of $\sim$3.
This is because the ``particle-particle" and ``hole-hole" components
such as [301 3/2]$\otimes$[422 5/2] or
[310 1/2]$\otimes$[431 3/2] have much less effect. A similar
quenching is calculated  for the $0^-$ state, which  becomes
a pure particle-hole
excitation  if pairing is reduced. On the other hand, the
characteristics of the  2$^-$ state are only
weakly influenced by  pairing.

The lowest $K^\pi$=0$^-$ excitations in prolate configurations of
$^{78,80,82}$Sr carry a rather
  weak collectivity. Like in $^{76}$Sr,
 in the
``weak pairing" variant
those states become almost pure particle-hole excitations.
 A similar situation is predicted for
the $K^\pi$=$1^-$ and $3^-$ states. The $K^\pi$=2$^-$
modes are found to be slightly more collective
compared to other modes with $K$=0, 1, and 3. They are expected to appear
at about $E_{ex}$=2.7 MeV and they carry  $E3$ strength
around 6 s.p.u. On the other hand, if pairing is reduced those states
become less collective.

The most collective
octupole excitations in the oblate configuration of
$^{82}$Sr are the $K^\pi$=1$^-$ and 2$^-$ states
[$E_{ex}$$\sim$2.7 MeV, $B(E3)$$\sim$7 s.p.u.].
The calculations also predict a low-lying weakly-collective
$K^\pi$=1$^-$ excitation in the superdeformed configuration of $^{82}$Sr
($E_{ex}$$\sim$2.3 MeV, $B(E3)$$\sim$10 s.p.u.).

Figures 12 and 13 display calculated
low-lying negative
parity states built upon the oblate and prolate configurations in the Kr
isotopes, respectively.
On the average, negative parity states
in Kr's are slightly more collective than those in Sr's.
The $K^\pi$=0$^-$ prolate excitations
are almost
pure two-quasiparticle states.
 The
$K^\pi$=1$^-$ states
and the $K^\pi$=2$^-$ oblate states
 resemble octupole vibrations; they have
$E_{ex}$$\sim$2.5 MeV, $B(E3)$$\sim$7 s.p.u.
The most collective octupole state in the Kr isotopes is the
 $K^\pi$=3$^-$ excitation
($E_{ex}$$\sim$3.2 MeV, $B(E3)$$\sim$10 s.p.u.)
  in $^{72}$Kr built
upon the oblate minimum. However, when  pairing is reduced this state
becomes  almost  a pure particle-hole excitation.

Finally, the results for the Zr isotopes are shown in Fig.\ 14.
The lowest negative-parity excitations in $^{80}$Zr and
$^{82}$Zr (prolate configuration) have a two-quasiparticle character.
The $K^\pi$=0$^-$, 1$^-$, and 2$^-$ modes in the oblate minimum
of $^{82}$Zr are weakly collective, with $B(E3)$$\sim$5--9 s.p.u.
Interestingly, the $B(E3)$ rates for these states do not depend strongly
on pairing. This is because their dominant  two-quasiparticle
components are the particle-like $(g_{9/2})_{1/2,3/2,5/2}$
 orbitals and the  hole-like negative-parity $p_{3/2}$$\oplus$$f_{5/2}$
levels
with $\Omega$=1/2 and 3/2.

\section{Conclusions}

In the light zirconium region
there are many excellent candidates for the low-lying
1$^+$ states with unusually large $B(M1;0^+\rightarrow 1^+)$ rates,
around 1--2 $\mu_N^2$. The best prospects are the Z=N nuclei,
such as $^{76}$Sr (prolate), $^{80}$Zr (prolate), and
$^{72}$Kr (oblate), where protons and neutrons contribute
equally strongly to the $M1$  collectivity.
 Interestingly, the unusually strong low-energy
$M1$ strength in those nuclei has a simple interpretation
in terms of $(g_{9/2})^2$ excitations, i.e., it does not result
from a simplistic scissors mode. Also, it does
not resemble the strong $M1$ transitions
known in the light Z=N nuclei \cite{[Kur63]}, mainly of the
spin-flip origin.

In $^{76}$Sr and neighboring nuclei, the $1^+$ excitations are predicted
to appear just above the $\pi$=-- intrinsic states.
Generally, the
$K^\pi$=0$^-$, 1$^-$, 2$^-$, and 3$^-$ bandheads are calculated
to be very weakly collective in well-deformed
proton-rich Kr, Sr, and Zr nuclei
(except maybe $^{72}$Kr). Namely, the low-lying negative-parity
states have a dominant two-quasiparticle character
when they are built on an intrinsic state
with a  large quadrupole deformation.
There is no clear correlation between the excitation energy of the
3$^-$ state and the magnitude of the $B(E3)$$\uparrow$ value
in the nuclei from the proton-rich Sr-Zr region.

The results of our calculations are quite sensitive
to the strength of  pairing interaction. In general, the weaker
the pairing
correlations, the more (less) collective are
the $M1$ ($E3$) excitations. There exists some indirect experimental
evidence supported by calculations,
see Sec. \ref{pairing}, that pairing is seriously reduced in
some excited states of well-deformed
nuclei from the A$\sim$80 mass region.
We hope that future measurements of excited states
in the well deformed nuclei around $^{76}$Sr, especially their lifetimes,
will shed new light on the collectivity of $M1$ and $E3$ states
and, indirectly, on the magnitude of pairing correlations in this
mass region.

\acknowledgements
The Joint Institute for Heavy Ion
Research has as member institutions the University of Tennessee,
Vanderbilt University, and the Oak Ridge National Laboratory; it
is supported by the members and by the Department of Energy
through Contract Number DE-FG05-87ER40361 with the University
of Tennessee.  Theoretical nuclear physics research at the
University of Tennessee is supported by the U.S. Department of
Energy through Contract Number DE-FG05-93ER40770.
This work was also partly supported by the
Yamada Science Foundation, Osaka, Japan,  and
the Swedish Natural Science Research Council.
The calculations were  supported in part by RCNP, Osaka
University, as RCNP Computational Nuclear Physics Project (Project No.
93-B-02).

\clearpage

\begin{center}
{\Large Figure captions}
\end{center}

\bigskip
\noindent
{\bf Figure 1:}
Neutron single-particle levels in $^{78}$Kr as functions of the
quadrupole deformation $\beta_2$ ($\beta_4$=0). The Nilsson states
are labelled by means of the asymptotic quantum numbers,
[$N n_z \Lambda~\Omega$].

\bigskip
\noindent
{\bf Figure 2:}
Predicted excitation energies of low-lying 1$^+$ states of prolate
configurations in $^{76,78,80,82}$Sr,
oblate minimum in $^{82}$Sr [82(o)], and the superdeformed configuration
in  $^{82}$Sr [82(SD)]. The numbers indicate
the $B(M1;g.s.\rightarrow 1^+)$ values
(in $\mu_N^2$) for transitions greater than 0.5~$\mu_N^2$.
Only states with
$B(M1;g.s.\rightarrow 1^+)$$>$0.1~$\mu_N^2$ are shown
(solid lines: $B(M1)$$>$0.3~$\mu_N^2$,
dashed lines: $B(M1)$$<$0.3~$\mu_N^2$).
 The upper portion shows
the results obtained with standard pairing, $\Delta_{std}$, see
Table~\protect\ref{defs}.  The results obtained with
pairing reduced by 50\% are displayed
in the lower portion.

\bigskip
\noindent
{\bf Figure 3:}
$B(M1;g.s.,K^\pi$=$0^+\rightarrow K^\pi=1^+)$
values for $^{76}$Sr calculated in
{\sc rpa} as a
function of the excitation energies of $1^+$ states. The  summed values
per 1  MeV energy bin
are plotted as a histogram (solid lines).  For
reference, the $B(M1)$ values associated with   spin   part only
($g_l$=0, dotted line) or  orbital   part
only ($g_s$=0, dashed line) are also shown.
The $g$-factors used are  $g_l$=$g_l^{free}$
and $g_s$=$(0.85)g_s^{free}$.
The upper (lower)  diagram represents the ``standard pairing"
(``weak pairing") variant of the calculations.

\bigskip
\noindent
{\bf Figure 4:}
Similar to Fig. 2 (standard pairing) but for
the Kr
isotopes.

\bigskip
\noindent
{\bf Figure 5:}
Similar to Fig.\ 2 (standard pairing) but for
the Zr
isotopes.

\bigskip
\noindent
{\bf Figure 6:}
The lowest 3$^-$ energy level (in keV), observed experimentally
for doubly even nuclei from the light zirconium region.
The dashed lines represent the lowest contours, at 2.2 and 2.3\ MeV.

\bigskip
\noindent
{\bf Figure 7:}
Predicted excitation energies of low-lying intrinsic
$K^\pi$=0$^-$, 1$^-$, 2$^-$, and 3$^-$ states
in $^{76,78,80,82}$Sr. The numbers indicate
the $B(E3;g.s.\rightarrow K^-)$ values in s.p.u.
[1 s.p.u.=0.416\,10$^{-6}$A$^2e^2b^3$, cf.
ref.\ \cite{[Spe89]}].
They are shown for the states
with $B(E3)$$>$1\,s.p.u. Other states represent non-collective
$\pi$=-- excitations. The solid lines correspond to states
with $B(E3)$$>$3\,s.p.u. while the dashed lines
correspond to states
with $B(E3)$$<$3s.p.u.
The results were obtained with standard pairing, $\Delta_{std}$, see
Table~\protect\ref{defs}.

\bigskip
\noindent
{\bf Figure 8:}
Absolute values of forward {\sc rpa}
amplitudes of the lowest  $K^\pi$=0$^-$ states
built upon prolate minima in the Sr
isotopes versus the quasiparticle configuration (numbered according
to their excitation energies) for neutrons
(solid lines) and protons (dashed lines).
All amplitudes whose absolute values greater than $5\times 10^{-2}$ are
indicated.
(Note that due to the time-reversal symmetry
each amplitude contributes to the
intrinsic wave function
twice.)
The results were obtained with standard pairing, $\Delta_{std}$, see
Table~\protect\ref{defs}.

\bigskip
\noindent
{\bf Figure 9:}
Similar to Fig.\ 8  but for
the lowest $K^\pi$=1$^-$ states in the Sr
isotopes.

\bigskip
\noindent
{\bf Figure 10:}
Similar to Fig.\ 9
but for
the lowest $K^\pi$=2$^-$ states in the Sr
isotopes.

\bigskip
\noindent
{\bf Figure 11:}
Similar to Fig.\ 9 but for
the lowest $K^\pi$=3$^-$ states in the Sr
isotopes.

\bigskip
\noindent
{\bf Figure 12:}
Similar to Fig.\ 7 but for
the lowest $\pi$=--  states in oblate configurations in the
$^{72,74,76}$Kr
isotopes.

\bigskip
\noindent
{\bf Figure 13:}
Similar to Fig.\ 7 but for
the lowest $\pi$=--  states in prolate configurations in the
$^{72,74,76}$Kr
isotopes.

\bigskip
\noindent
{\bf Figure 14:}
Similar to Fig.\ 7 but for
the lowest $\pi$=--  states in the $^{80,82}$Zr
isotopes.

\newpage
\begin{table}
\begin{center}
\caption{
Calculated equilibrium  shape
deformations $\beta_2$ and $\beta_4$, and
proton and neutron pairing gaps, $\Delta_p$ and $\Delta_n$ (in MeV),
at selected oblate and prolate configurations of Kr, Sr and Zr isotopes.
According to calculations, the oblate $I$=0 minima lie lower in energy
than the prolate $I$=0 minima in $^{72,74,78}$Kr, $^{82}$Sr, and $^{82}$Zr.
For $^{82}$Sr the calculations were also performed at superdeformed
configuration with $\beta_2$=0.45.
}\label{defs}
\vspace{1cm}
\begin{tabular}{cc|rrcc|rrcc}
\multicolumn{2}{c|}{Nucleus} & \multicolumn{4}{c|}{Oblate}
& \multicolumn{4}{c}{Prolate}\\[1mm]
\hline
Z & N & $\beta_2$ & $\beta_4$ & $\Delta_p$ & $\Delta_n$ &
$\beta_2$ & $\beta_4$ & $\Delta_p$ & $\Delta_n$ \\[1mm]
 \hline
36 & 36 & --0.31 & --0.010 & 1.34 & 1.23 & 0.35 & 0.016 & 1.40 & 1.31\\
   & 38 & --0.30 & --0.016 & 1.26 & 1.46 & 0.37 & 0.0 & 1.31 & 1.12\\
   & 40 & --0.25 & --0.036 & 1.32 & 1.54 & 0.36 & --0.016 & 1.24 & 1.25\\
   & 42 & --0.24 & --0.050 & 1.28 & 1.48 & 0.32 & --0.023 & 1.18 & 1.46\\
   & 44 & --0.23 & --0.050 & 1.24 & 1.46 &      &        &      &     \\
   &    &        &         &      &      &      &        &      &     \\[-3mm]
38 & 38 &  &  &  &  & 0.39 & --0.016 & 1.14 & 0.99  \\
   & 40 &  &  &  &  & 0.39 & --0.029 & 1.01 & 1.04\\
   & 42 &  &  &  &  & 0.37 & --0.030 & 0.93 & 1.34\\
   & 44 & --0.22 & --0.065  & 1.35  & 1.37  & 0.28 & --0.020 & 1.15 & 1.48\\
   & 44 &  &  &  &  & 0.45 & 0.0 & 0.83 & 1.45\\
   &    &        &         &      &      &      &        &      &     \\[-3mm]
40 & 40 &  &  &   &   & 0.40 & --0.037 & 1.06 & 0.88 \\
   & 42 & --0.22 & --0.078 & 1.39 & 1.31 & 0.39 & --0.038 & 0.96 & 1.26 \\[1mm]
\hline
\end{tabular}
\end{center}
\end{table}

\end{document}